\documentclass[sigconf]{acmart}

\usepackage[ruled]{algorithm2e}
\usepackage{multirow}
\usepackage{pifont}
\usepackage{balance}

\AtBeginDocument{%
  }

\copyrightyear{2025}
\acmYear{2025}
\setcopyright{acmlicensed}
\acmConference[WWW Companion '25]{Companion Proceedings of the ACM Web Conference 2025}{April 28-May 2, 2025}{Sydney, NSW, Australia}
\acmBooktitle{Companion Proceedings of the ACM Web Conference 2025 (WWW Companion '25), April 28-May 2, 2025, Sydney, NSW, Australia}
\acmDOI{10.1145/3701716.3715231}
\acmISBN{979-8-4007-1331-6/25/04}

\begin{document}

\title{HCMRM: A High-Consistency Multimodal Relevance Model for Search Ads}


\author{Guobing Gan}
\affiliation{%
  \institution{Kuaishou Technology}
  \city{Beijing}
  \country{China}}
\email{ganguobing@kuaishou.com}

\author{Kaiming Gao}
\affiliation{%
  \institution{Kuaishou Technology}
  \city{Beijing}
  \country{China}}
\email{gaokaiming@kuaishou.com}

\author{Li Wang}
\affiliation{%
  \institution{Kuaishou Technology}
  \city{Beijing}
  \country{China}}
\email{wangli18@kuaishou.com}

\author{Shen Jiang}
\affiliation{%
  \institution{Kuaishou Technology}
  \city{Beijing}
  \country{China}}
\email{jiangshen@kuaishou.com}

\author{Peng Jiang}
\affiliation{%
  \institution{Kuaishou Technology}
  \city{Beijing}
  \country{China}}
\email{jiangpeng@kuaishou.com}

\renewcommand{\shortauthors}{Guobing Gan, Kaiming Gao, Li Wang, Shen Jiang, and Peng Jiang}

\begin{abstract}
Search advertising is essential for merchants to reach the target users on short video platforms. Short video ads aligned with user search intents are displayed through relevance matching and bid ranking mechanisms. This paper focuses on improving query-to-video relevance matching to enhance the effectiveness of ranking in ad systems. Recent vision-language pre-training models have demonstrated promise in various multimodal tasks. However, their contribution to downstream query-video relevance tasks is limited, as the alignment between the \textbf{pair} of visual signals and text differs from the modeling of the \textbf{triplet} of the query, visual signals, and video text. In addition, our previous relevance model provides limited ranking capabilities, largely due to the discrepancy between the binary cross-entropy fine-tuning objective and the ranking objective.
To address these limitations, we design a \textbf{h}igh-\textbf{c}onsistency \textbf{m}ultimodal \textbf{r}elevance \textbf{m}odel (\textbf{HCMRM}). It utilizes a simple yet effective method to enhance the consistency between pre-training and relevance tasks. Specifically, during the pre-training phase, along with aligning visual signals and video text, several keywords are extracted from the video text as \textbf{pseudo-queries} to perform the triplet relevance modeling. For the fine-tuning phase, we introduce a \textbf{hierarchical softmax} loss, which enables the model to learn the order within labels while maximizing the distinction between positive and negative samples. This promotes the fusion ranking of relevance and bidding in the subsequent ranking stage.
The proposed method has been deployed in the Kuaishou search advertising system for over a year, contributing to a 6.1\% reduction in the proportion of irrelevant ads and a 1.4\% increase in ad revenue.
\end{abstract}




\maketitle

\section{Introduction}

\begin{figure*}
  \includegraphics[width=\textwidth]{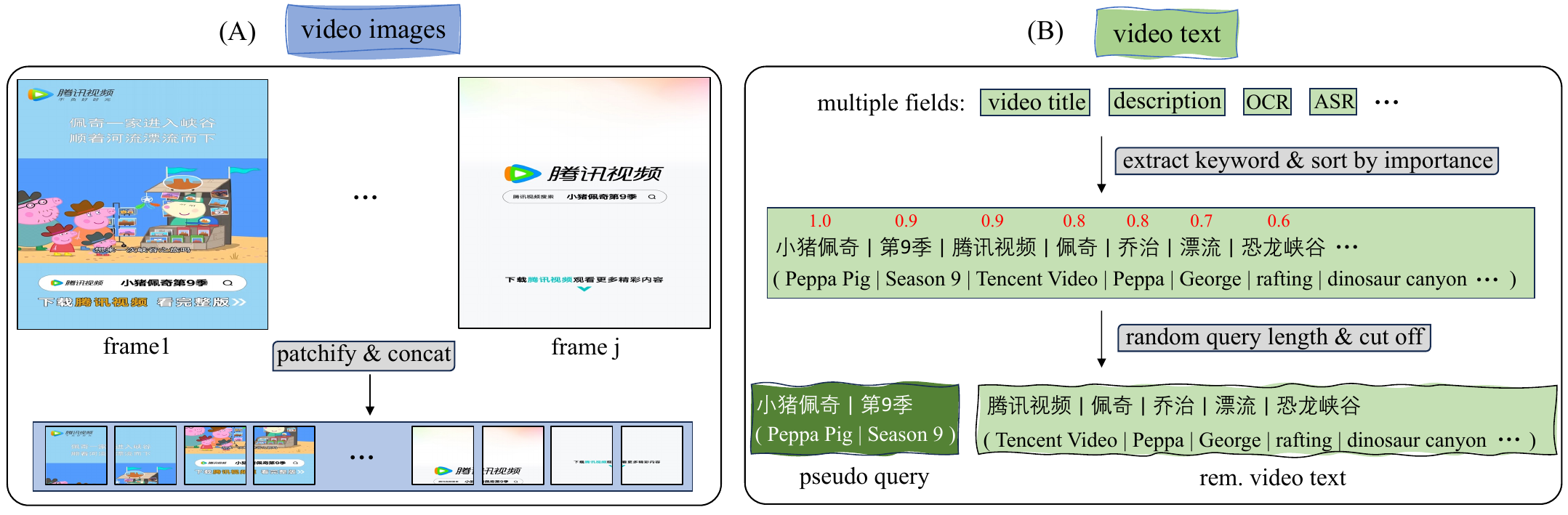}
  \caption{An example of the short video ad, including both video signals and text information. (A) is a diagram of the video image processing procedure. The uniformly sampled frames are blocked and spliced before being input into the model. The upper part of (B) shows that the video text consists of keywords extracted from multiple text fields in the video, arranged in descending order of importance. The lower part of (B) illustrates the process of generating pseudo-queries in the pre-training.}
  \Description{An example of the short video ad, including both video signals and text information.}
  \label{fig:video_introduce}
\end{figure*}

The consumption of short videos has become popular in our daily lives. Hundreds of millions of users visit short video platforms (e.g., Likee, Instagram Reels, TikTok, and Kuaishou) every day \cite{Xiao2023ExploringTF}. Search advertising is one of the primary ad types on short video platforms, generating substantial commercial prospects. It displays short video ads that align with user search intent and possess high commercial value through relevance matching and bidding ranking mechanisms. The relevance module is crucial in the ad system, as it assesses the degree of relevance between a short video ad and a given query to decide whether the ad should be filtered out. Additionally, relevance scores from this module and bid scores from the bidding module are combined to determine the ad ranking.

To calculate the relevance between a query and a short video ad, it is necessary to integrate the visual signals of the video with its textual information (incl., video title, OCR, ASR, etc). In recent years, vision-language pre-training has achieved remarkable success in various multimodal tasks \cite{Li2023MultimodalFM,Zhang2024MMLLMsRA}. However, most approaches focus on aligning visual signals with video or image descriptions. This pairwise alignment differs from the relevance modeling of the triplet consisting of the query, visual signals, and video text. Consequently, traditional pre-training methods have limited effectiveness for query-video relevance tasks. To enhance query awareness in pre-training, some works \cite{Ye2023QueryawareMB,Zhu2023QueryLIFEQL} in the fields of e-commerce search and long video search suggest leveraging post-click behavioral data for pre-training. However, click data is usually noisy due to exposure bias and clickbait \cite{Zou2021PretrainedLM}. 
Moreover, most videos have few impressions or click throughs due to the long-tail distribution, resulting in a limited amount of query-video pairs with high click-through rates and confidence levels available for pre-training.
Furthermore, these query-aware methods \cite{Ye2023QueryawareMB,Zhu2023QueryLIFEQL} typically require the introduction of an additional query tower built on mainstream multimodal models like ALBEF \cite{Li2021AlignBF} and BLIP \cite{Li2022BLIPBL}, resulting in substantial modification costs and increases the overall complexity of the architecture.

In the manually annotated data for query-video ad pairs, relevance levels are categorized into four tiers: Bad, Less, Good, and Excellent. To maximize the distinction between positive and negative samples, our previous relevance model employed binary cross-entropy loss (i.e., Bad and Less refer to 0, Good and Excellent refer to 1) for fine-tuning. This approach helps the system better filter out irrelevant query-video ad pair but fails to account for the ordinal nature of the labels, which weakens the ranking ability of relevance models in the subsequent ad ranking stage.

To address these challenges, we propose the high-consistency multimodal relevance model (HCMRM) which aims to enhance the consistency of the pre-training and downstream query-video relevance tasks, as well as the coherence between the relevance module and the ad ranking module.
To improve the computational efficiency, we convert lengthy short video texts into a keyword sequence, with the keywords arranged in descending order of their importance weights, as shown in the upper part of Figure~\ref{fig:video_introduce}(B).
For the query-video relevance model, we leverage the architecture of ALBEF \cite{Li2021AlignBF} with minimal modification cost. This architecture includes a text encoder, a vision encoder, and a fusion encoder, as shown in Figure~\ref{fig:pretrained_model}. 
During the pre-training phase, common tasks such as contrastive learning, image-text matching, and masked language modeling are employed to align video images (frames) with video texts (keywords). Additionally, several relatively important keywords on the left side of the video keyword sequence are extracted and used as a pseudo-query, while the remaining portion continues to serve as the video text. The pseudo-query and its corresponding video form a positive sample, while it is paired with other videos as negative samples. A hard negative sampling mechanism is also employed, similar to that used in image-text matching tasks \cite{Li2021AlignBF}. Therefore, using this pseudo-query-video matching mechanism, we can perform training consistent with the downstream relevance task in the pre-training phase.
For fine-tuning using human-labeled data, we introduce a symmetric hierarchical softmax loss based on the symmetry of the four-level relevance labels. This relevance loss function consists of three binary cross-entropy objectives: one for distinguishing between positive and negative samples, one for distinguishing between Good and Excellent within the positive samples, and one for distinguishing between Bad and Less within the negative samples. This approach enables the model to maximize the distinction between positive and negative samples while learning the ordinal relationships among the labels, thereby strengthening the ranking capability of the relevance model.

To validate the benefits of the proposed HCMRM, we have conducted extensive experiments both offline and online. The experimental results show that HCMRM achieves significant improvements in metrics such as AUC and Spearman's rank correlation coefficient compared to our previous relevance model and various baseline models. Furthermore, it achieves a 6.1\% reduction in the proportion of irrelevant ads and a $1.4\%$ increase in ad revenue for Kuaishou's search advertising. Our main contributions can be summarized as follows:
\begin{itemize}
\item This is one of the earliest works on query-short video multimodal relevance, particularly in the advertising domain.
\item We propose a multimodal pre-training framework that enhances the consistency between pre-training and downstream relevance tasks, building upon mainstream multimodal models without significant modification costs.
\item We introduce a symmetric hierarchical softmax loss that enhances the ranking capability of relevance models, optimizing the ranking process for advertising systems.
\end{itemize}

\section{Related Work}

\subsection{Vision-Language Pre-training}
Vision-language pre-training has recently achieved remarkable success in various multimodal tasks by learning multimodal representations on large-scale image-text or video-text pairs \cite{Radford2021LearningTV, Fang2022EVAET, Li2021AlignBF, Bain2021FrozenIT, Luo2021CLIP4ClipAE, Mu2021SLIPSM, Ma2022XCLIPEM, Huang2022CloverTA, Xu2021VideoCLIPCP, Chen2023InternVS, Xu2023mPLUG2AM, Liu2023VisualIT, li2024llava}. 
Methods like CLIP \cite{Radford2021LearningTV}, ALIGN \cite{Jia2021ScalingUV}, and DeCLIP \cite{Li2021SupervisionEE} leverage contrastive learning for pre-training on large massive noisy web data, achieving promising performance on cross-modal retrieval tasks. 
ALBEF \cite{Li2021AlignBF} and FLAVA \cite{Singh2021FLAVAAF} introduce multimodal encoders to perform complex cross-modal interactions, leading to superior performance in multimodal reasoning tasks. SimVLM \cite{Wang2021SimVLMSV}, CoCa \cite{Yu2022CoCaCC}, and BLIP \cite{Li2022BLIPBL} add multimodal decoders, enabling new capabilities in image-conditioned text generation. Recent multimodal large models like BLIP-2 \cite{Li2023BLIP2BL}, LLAVA \cite{li2024llava}, and Qwen-VL  \cite{Bai2023QwenVLAV} harness large language models as a cognitive powerhouse, further advancing performance in various multimodal tasks \cite{Zhang2024MMLLMsRA, Yin2023ASO}.

\subsection{Multimodal Models for Relevance}
Early relevance methods focus on the text modality \cite{Yao2021LearningAP, Chang2021ExtremeML, Zou2021PretrainedLM, Liu2021Que2SearchFA}. In recent years, some multimodal relevance methods have emerged. For instance, Query-LIFE \cite{Zhu2023QueryLIFEQL} utilizes a query-based multimodal fusion to effectively incorporate the image and title of products, improving relevance and conversion efficiency in e-commerce search. QUALITY \cite{Ye2023QueryawareMB} introduces a query-aware multimodal model to enhance the ranking relevance of video search. However, these methods necessitate the addition of extra query towers, which complicates both the model structure and the pre-training process. In contrast, our HCMRM can directly reuse existing mainstream multimodal models. Additionally, their pre-training processes rely on constructing query-item pairs from the online click logs. However, click data is often noisy and reliable query-item pairs are typically limited. This limits the effectiveness of the pre-training.

\subsection{Relevance Fine-tuning Loss}
In industrial scenarios, the relevance of query-item pairs is manually divided into multiple levels \cite{Yao2021LearningAP, Zou2021PretrainedLM, Wen2023EnhancingDI, Ye2023QueryawareMB}. For instance, Baidu search ads categorize query-image relevance as \{0, 1, 2\}, with 0 as negative and 1 or 2 as positive, and binary cross-entropy loss is used for relevance fine-tuning training \cite{Wen2023EnhancingDI}. Tencent Video categorizes query-video relevance into four levels \cite{Ye2023QueryawareMB}. They utilize ordinal regression loss for training, which is a good approach to account for the ordinal nature of relevance labels.

\begin{figure*}
  \includegraphics[width=0.95\textwidth]{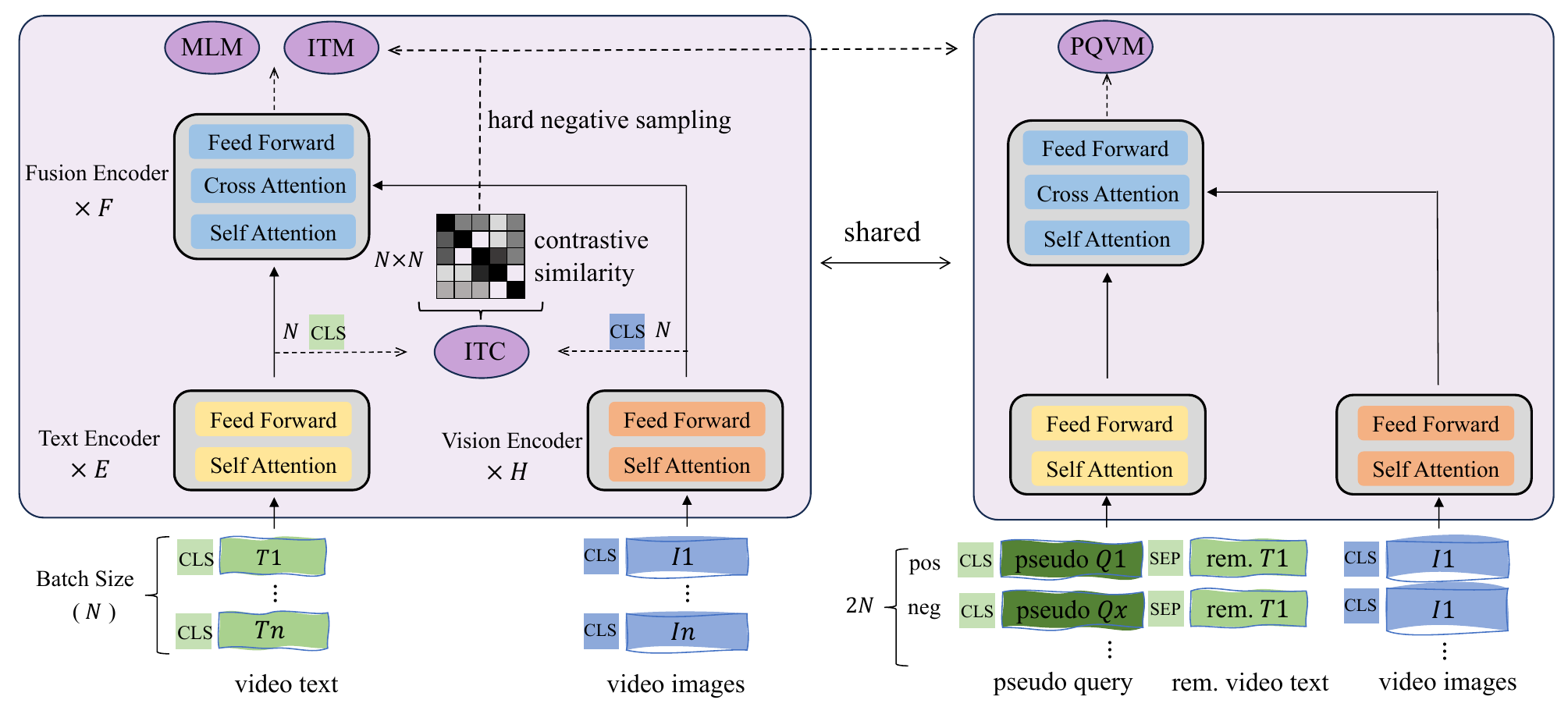}
  \caption{Overview of HCMRM. It is built based on ALBEF with minimal modifications and pre-trained using four objectives: Image-Text Contrastive Learning (ITC), Image-Text Matching (ITM), Masked Language Modeling (MLM), and Pseudo-Query-Video Matching (PQVM). Note that the downstream relevance task between query and short video ad is consistent with PQVM.}
  \Description{Overview of the proposed multimodal relevance model, HCMRM.}
  \label{fig:pretrained_model}
\end{figure*}

\section{Methodology}

\subsection{Preliminary}

We first introduce the role of the relevance module in the Kuaishou search advertising system and define the core relevance problem. 

\textbf{Role of the Relevance Module.} The relevance module plays a \verb|filtering| and \verb|ranking| role in the search advertising system of short videos. Specifically, it receives a user's query and a set of candidate short video ads retrieved by front-end modules like recall. The relevance module scores the relevance between the query and each short video. On the one hand, different ad slots have specific relevance thresholds, and only ads with relevance scores exceeding these thresholds have the opportunity to be delivered to the corresponding ad slots. On the other hand, during the ad ranking stage, relevance scores are combined with bid scores from the bidding module to determine the final ranking of the ads.

\textbf{Definition of Relevance Problem.}
The relevance problem of query-video matching (QVM) can be conceptualized as a scoring function
$\mathcal{F}$. This function takes two inputs: a user-issued query ($Q$) and a short video ad ($V$) triggered by the system. The output is a relevance score ($r$) that quantifies the degree of relevance between the query and the ad. 
Since a video contains visual signals (video frames/images) and textual information (i.e., video titles, descriptions, OCR, ASR, etc.), this relevance problem can be formulated as a function involving a \verb|triplet|. It is defined as follows:
\begin{equation}
\begin{aligned}
r = \mathcal{F}(Q, V) = \mathcal{F}(Q, I, T) 
\end{aligned}
\end{equation}
where $I$ and $T$ are the visual signals and text information of the given short video ad, respectively.

\subsection{Feature Engineering}

\begin{algorithm}
 \SetAlgoNoLine  
 \caption{Keyword Extraction for Video Text}
  \KwIn{\\
    \quad A list of $l$ text fields from a video ad: $f=[f_1, \cdots, f_l]$;\\
    \quad The predefined weights for each field: $\alpha=[\alpha_1, \cdots, \alpha_l]$;\\
    \quad The maximum number of generated keywords: $k$;\\}
  \KwOut{\\
    \quad The generated keyword sequence: $t$;\\}
    Initialize an empty dictionary: $d=\{word:importance\}$; \\
    \For{$i=1$ to $l$}
    {
        $W = Tokenize(f_i)$; \\
        \For{each $w \in W$}
        {
            $\omega = Importance(w, f_i)$; \\
            $\omega = \omega * \alpha_i$; \\
            $d[w] = d.get(w, 0) + \omega$; \\
        }
    }
    $top\_words =sorted$($d$, key=$d$.get, reverse=True)$[:k]$; \\
    $t = "|".join(top\_words)$; \\
    return $t$; \\
 \label{keyword_extraction}
\end{algorithm}

We introduce the pre-processing of short video ads composed of visual and textual information as follows.

\textbf{Visual Signals.} For the visual signals of a video, $J$ frames are extracted from the video at equal time intervals. These frames are then divided into patches, which are concatenated in chronological order to form a patch sequence for input into the model, as illustrated in Figure~\ref{fig:video_introduce}(A).

\textbf{Textual Information.} The video text contains multiple fields such as title, description, OCR, and ASR, making it very lengthy and redundant. We employ keyword extraction to the video text to improve computational efficiency, as shown in Figure~\ref{fig:video_introduce}(B). The keyword extraction algorithm, detailed in Algorithm~\ref{keyword_extraction}, aims to extract the top $k$ most important words from multiple text fields of a video. Specifically, each text field is assigned a predefined weight $\alpha$. Firstly, initialize an empty dictionary where the keys are words and the values are their importance. Next, perform tokenization on each text field. Then, calculate the importance of each tokenized word. The importance assessment method is based on TF-IDF and named entity recognition. The importance of each word is weighted by the importance of the corresponding text field and summed accordingly. Finally, all words are sorted in descending order of importance, and the top $k$ words are connected to form a sequence of keywords.

\subsection{Model Architecture}
As illustrated in Figure~\ref{fig:pretrained_model}, HCMRM largely reuses the structure of ALBEF \cite{Li2021AlignBF}, consisting of a text encoder, an image encoder, and a multimodal fusion encoder. The image encoder is initialized using the first 6 layers of ViT-B/32 \cite{Dosovitskiy2020AnII} \footnote{ViT's position embeddings are extended to support long patch sequences of videos.}. The image patch sequence from video frames is used as input, resulting in an image feature sequence $[i_{cls}, i_1, \cdots, i_K]$, where $i_{cls}$ is the embedding of the token \verb|[CLS]|. The text encoder is initialized using the first 6 layers of BERT-base-chinese \cite{Devlin2019BERTPO}, and the multimodal fusion encoder is initialized using the last 6 layers of BERT-base-chinese. The text encoder transforms an input text $T$ into a sequence of
embeddings $[t_{cls}, t_1, \cdots, t_L]$. The multimodal fusion encoder takes the image and text features output by the previous two encoders as input and fuses them through the cross-attention mechanism at each of its sub-layers.

\subsection{Pre-training}
HCMRM is pre-trained using the following objectives: Image-Text Contrastive Learning (\textbf{ITC}), Image-Text Matching (\textbf{ITM}), Masked Language Modeling (\textbf{MLM}), and Pseudo-Query-Video Matching (\textbf{PQVM}), as shown in Figure~\ref{fig:pretrained_model}. Online contrastive hard negative mining based on ITC is employed to enhance the performance of ITM and PQVM.

\textbf{Image-Text Contrastive Learning.} ITC aims to align the visual signals and text information of videos before fusion. Based on widely used contrastive learning methods \cite{Radford2021LearningTV, Li2021AlignBF, Li2022BLIPBL, Yu2022CoCaCC}, ITC encourages positive image-text pairs to have similar representations in contrast to negative pairs. We follow the ITC loss of ALBEF \cite{Li2021AlignBF} and BLIP \cite{Li2022BLIPBL}, the normalized \verb|[CLS]| embeddings from the image and text encoders serve as unimodal representations, respectively. Meanwhile, a momentum model is introduced to produce unimodal embeddings and soft image-text similarity targets. For further details on the ITC task, please refer to ALBEF \cite{Li2021AlignBF}.

\textbf{Image-Text Matching.} ITM predicts whether a pair of video images and text is positive (matched) or negative (not matched). The \verb|[CLS]| embedding output from the multimodal fusion encoder serves as a joint representation of the pair. It is projected into a two-dimensional space through a fully-connected layer and trained using binary cross-entropy loss. Similar to ALBEF \cite{Li2021AlignBF}, we utilize the contrastive similarity distribution from ITC to identify in-batch hard negatives. For each image in the mini-batch, text that is more similar to the image within the same batch is sampled with a higher probability to create a negative sample with the current image. Likewise, a hard negative image is also sampled for each text.

\textbf{Masked Language Modeling.} MLM utilizes both video images and video text to predict the masked textual tokens. Similar to BERT \cite{Devlin2019BERTPO}, the special token \verb|[MASK]| is used to randomly replace a portion of the tokens in the video text. However, unlike BERT, we apply masking at the word level rather than at the subword or character level to improve the model's ability to learn entities.

\textbf{Pseudo-Query-Video Matching.} PQVM predicts whether a triplet of query, video images, and video text is positive (matched) or negative (not matched), which is consistent with downstream relevance task. The difference is that the pre-training utilizes synthetic query-video pairs, while the downstream fine-tuning utilizes real human-annotated query-video pairs.

The key to the PQVM pre-training task is the synthesis of query-text-image triplets. Based on the assumption that a query should represent the key content of a short video, we propose a simple method for synthesizing pseudo-queries using video texts. \textit{Since the video text has already been processed into a sequence of keywords ranked by importance, the keywords at the beginning of the sequence can effectively serve as queries.} To simulate real search scenarios, the number of keywords used as pseudo-queries is a random number within the range [a, b] during training. The video text is divided into two parts based on this number, with the remaining portion still serving as the video text, as follows:
\begin{equation}
\begin{aligned}
inx &= \text{randint}(a, \ b) \\
pseudo\ Q_i, \ \ rem. \ T_i &= \text{split}(T_i, \ inx)
\end{aligned}
\label{eq:pseudo_query}
\end{equation}
where $i \in [1, N]$ and $N$ is the batch size. $inx$ is the index for cutting the video text and also the length of the pseudo query.

The online contrastive hard negative
mining is also utilized in PQVM. Let $\mathcal{S}$ be an $N \times N$ similarity matrix, representing the similarity distribution between any two videos in a mini-batch. For HCMRM, $\mathcal{S}$ is the average of the softmax-normalized image-to-text and text-to-image
similarity matrices in the ITC task. The origin of these two matrices refers to ALBEF \cite{Li2021AlignBF}. For the $i$-th video $\{T_i, I_i\}$, its hard negative item is sampled based on the similarity distribution $\mathcal{S}_{i}$. Specifically, a video that is more similar to the $i$-th video within the same batch is sampled with higher probability. 
\begin{equation}
\begin{aligned}
x &= \text{multinomial}(\mathcal{S}_{i}) \\
\text{neg}_i &= \{pseudo\  Q_x, \ rem. \ T_i, \ I_i \}, \ x \neq i \\
\text{pos}_i &= \{pseudo\  Q_i, \ rem. \ T_i, \ I_i \}
\end{aligned}
\label{eq:pseudo_sample}
\end{equation}
where multinomial represents a sampling operation from a discrete probability distribution vector, where the return value is the index of the vector. Additionally, $\text{neg}_i$ represents the synthesized $i$-th hard query-video relevance negative sample, while $\text{pos}_i$ is the synthesized $i$-th query-video relevance positive sample. Note that the video text used in PQVM pre-training excludes the pseudo-query portion, and using the complete video text would lead to an information leakage issue.

As illustrated in the right part of Figure~\ref{fig:pretrained_model}, the query and the video text are concatenated with the special tokens \verb|[CLS]| and \verb|[SEP]| and then fed into the text encoder. Meanwhile, the image patch sequence from the same video as the text is prefixed with \verb|[CLS]| and encoded by the vision encoder. 

\begin{equation}
\begin{aligned}
z_T &= \text{Text Encoder}\left(\text{concat}\left(\verb|[CLS]|, Q, \verb|[SEP]|, T\right)\right) \\
z_I &= \text{Vision Encoder}\left(\text{concat}\left(\verb|[CLS]|, I\right)\right) \\
z &= \text{Fusion Encoder}\left(z_T, z_I\right)
\end{aligned}
\label{eq:sample_encoding}
\end{equation}
The \verb|[CLS]| embedding $z_{cls}$ output from the multimodal fusion encoder is utilized as a joint representation of the triplet of query, video images, and video text. Similar to ITM tasks, $z_{cls}$ is projected into a two-dimensional space via a linear layer and trained using binary cross-entropy loss. The PQVM loss can be expressed as:
\begin{equation}
\begin{aligned}
\mathcal{L}_{\mathrm{PQVM}} = -\mathbb{E}_{(Q, T, I) \sim D} [log\{P(y(Q, T, I)|(Q, T, I)\}]
\end{aligned}
\end{equation}
where $D$ is a distribution of in-batch samples, and $y(Q, T, I) \in \{0, 1\}$ represents whether the query $Q$ and video $(T, I)$ are matched. $P(y(Q, T, I)|(Q, T, I)$ is the output of the $z_{cls}$ followed by a two-class linear classifier.

Finally, HCMRM is pre-trained using the four tasks mentioned above, as shown in the loss function in Equation~\ref{loss:loss_pretrain}.
\begin{equation}
\begin{aligned}
\mathcal{L}_{\mathrm{pretrain}} = \mathcal{L}_{\mathrm{ITC}} + \mathcal{L}_{\mathrm{ITM}} + \mathcal{L}_{\mathrm{MLM}} + \mathcal{L}_{\mathrm{PQVM}}
\end{aligned}
\label{loss:loss_pretrain}
\end{equation}

\subsection{Fine-tuning}

\begin{figure}[h]
  \includegraphics[width=0.9\linewidth]{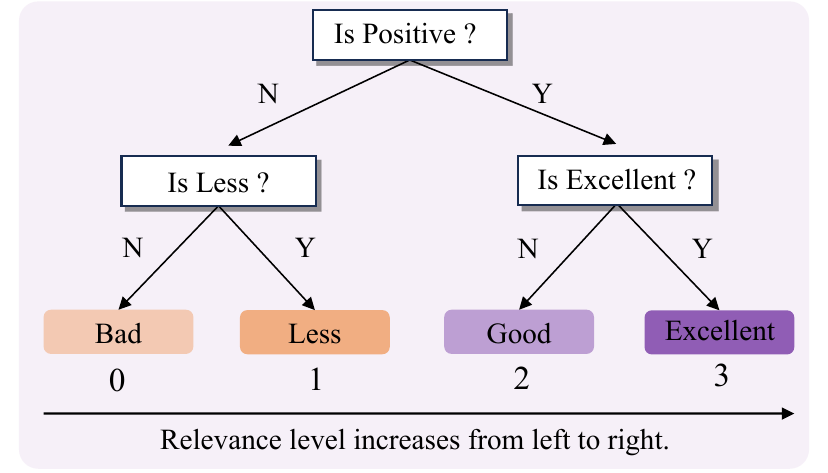}
  \caption{The hierarchical binary structure within the labels.}
  \Description{The hierarchical binary structure within the relevance labels.}
  \label{fig:ordinary_regression}
\end{figure}

For the downstream relevance task of query-video matching (QVM), HCMRM is fine-tuned on query-video ad data that is manually labeled into four relevance levels: \textbf{Bad (0)}, \textbf{Less (1)}, \textbf{Good (2)}, and \textbf{Excellent (3)}. It is evident that these four levels of relevance are ordinal labels and exhibit a balanced binary tree, as illustrated in Figure~\ref{fig:ordinary_regression}. Therefore, we intuitively model the query-video relevance task as a \textbf{two-layer symmetric hierarchical softmax} problem ~\cite{Tansey2018LeafSmoothedHS}. Specifically, three binary classifiers are used to predict the answers to the questions: \verb|'Is Positive?'|, \verb|'Is Less?'|, and \verb|'Is Excellent?'|, as shown in Equation~\ref{eq:hierarchical-softmax}.
\begin{equation}
\label{eq:hierarchical-softmax}
\begin{aligned}
 & p_{pos} = Sigmoid(z_{cls} W^{pos})  \\ 
 & p_{le} = Sigmoid(z_{cls} W^{le})  \\ 
 & p_{ex} = Sigmoid(z_{cls} W^{ex})
\end{aligned}
\end{equation}
where $W^{pos}$, $W^{le}$, and $W^{ex}$ are different linear projections that map the output \verb|[CLS]| embedding of HCMRM ($z_{cls}$) to a scalar. The probabilities for each label can be formulated as follows:
\begin{equation}
\label{eq:predicted-label}
\begin{aligned}
  p_0 &= (1-p_{pos})\times(1-p_{le}) \\
  p_1 &= (1-p_{pos})\times p_{le} \\
  p_2 &= p_{pos}\times(1-p_{ex}) \\
  p_3 &= p_{pos}\times p_{ex} \\
\end{aligned}
\end{equation}
where $p_0$, $p_1$, $p_2$, and $p_3$ represent the probabilities of Bad, Less, Good, and Excellent, respectively.
To obtain continuous relevance scores for ad ranking, the final relevance score is defined as the expectation of all labels, as formulated in Equation~\ref{eq:predicted-relevance-score}.
\begin{equation}
\label{eq:predicted-relevance-score}
\begin{aligned}
  r &= 0 \times p_0 + 1 \times p_1 + 2 \times p_2 + 3 \times p_3 \\
    &= p_1 + 2 \times p_2 + 3 \times p_3
\end{aligned}
\end{equation}
The relevance score 
$r$ ranges from [0, 3], with higher values indicating greater relevance.

The loss function for this hierarchical softmax problem is formalized as presented in Equation~\ref{eq:hierarchical-binary-loss}.
\begin{equation}
\label{eq:hierarchical-binary-loss}
\begin{aligned}
  \mathcal{L}_{\mathrm{QVM}} = -\frac{1}{N} \sum^{N}log(p_l) - \frac{1}{N} \sum^{N}(r - l)^2
\end{aligned}
\end{equation}
where $l \in \{0, 1, 2, 3\}$ represents the label. $p_l \in \{p_0, p_1, p_2, p_3\}$. This loss combines hierarchical softmax estimation with label regression.

\section{Experiments}

\subsection{Large-scale Industrial Datasets}
There are two datasets utilized for the pre-training and fine-tuning of HCMRM and baseline models. (1) The pre-training dataset consists of 200M Chinese short video materials from the Kuaishou platform, and each is a pair of video text and video images. Since the pre-training of HCMRM does not rely on real queries, we can utilize as much material as possible, making our pre-training dataset much larger than related work \cite{Zhu2023QueryLIFEQL, Ye2023QueryawareMB}.
(2) The fine-tuning dataset contains 2.5M <query, video> labeled data, and 230K samples are selected as the evaluation set. The labels categorize relevance levels from 0 to 3, corresponding to Bad, Less, Good, and Excellent.

\subsection{Evaluation Metrics}
We employ the following metrics for offline and online evaluations.

\textbf{Offline Metrics.} To comprehensively evaluate the model's ability to distinguish between positive and negative samples, as well as its sensitivity to different levels of relevance, we used three offline evaluation metrics: Area Under the Curve (\textbf{AUC}), Spearman's rank correlation coefficient (\textbf{Spearman}), and Pearson Product-Moment Correlation Coefficient (\textbf{Pearson}). For the AUC metric, relevance labels 0 and 1 are considered negative, while labels 2 and 3 are considered positive. The Spearman and Pearson metrics consider the partial order of the labels and assess the consistency between the prediction results and the ground truth.

\textbf{Online Metrics.} We conducted online A/B tests to compare HCMRM with our previous baseline model. We considered three metrics: Irrelevant Ratio, \#Conversions, and Ad Revenue. (1) \textbf{Irrelevant Ratio} refers to the proportion of cases within a sampled dataset that are labeled as "Bad" or "Less" by human annotators, calculated as $(\#\mathrm{Bad} + \#\mathrm{Less}) / (\#\mathrm{Bad} + \#\mathrm{Less} + \#\mathrm{Good} + \#\mathrm{Excellent})$. (2) An advertising conversion occurs when a potential customer views a video ad and subsequently takes an action deemed valuable to the advertiser's business, such as making an online purchase or calling the business from a mobile phone. \textbf{\#Conversions} refers to the total number of advertising conversions. (3) \textbf{Ad Revenue} refers to the revenue of the advertising system after assisting advertisers in achieving conversions.

\subsection{Baselines}
\label{baselines}
We compared HCMRM with the following models, encompassing unimodal, multimodal, and large models. 

\textbf{BERT.} This is a 12-layer BERT that is pre-trained on video text data using the same word-level masked language modeling task as HCMRM. Additionally, the processing of the video text follows the same approach as in HCMRM. The BERT model is initialized using BERT-base-chinese \cite{Devlin2019BERTPO}.

\textbf{BERT-VL.} This model is the previous version of the relevance module in our ad system. Its structure is identical to BERT \cite{Devlin2019BERTPO}, except that $J$ frames of video images are encoded into $J$ embeddings by MobileNetV2 \cite{Sandler2018MobileNetV2IR} and then concatenated with text embeddings before being input into the BERT model. For pre-training, we employ the word-level masked language modeling task, where a masked text token is predicted based on the combined representation of the visible tokens and visual features.

\textbf{ViLT.} We replicated ViLT \cite{Kim2021ViLTVT} in the short video advertising scenario. It is a vison-and-language Transformer without convolution or region supervision. The ViLT model is initialized
using ViT-B/32 \cite{Dosovitskiy2020AnII}. The image-text matching and word-level masked language modeling tasks are used for pre-training. 

\textbf{ALBEF.} ALBEF can be regarded as a version of HCMRM without the PQVM mechanism, as HCMRM is built on ALBEF \cite{Li2021AlignBF}. The image-text contrastive learning, image-text matching, and masked language modeling tasks are used for pre-training. Unlike the vanilla ALBEF \cite{Li2021AlignBF}, word-level rather than subword-level masked language modeling is used.

\textbf{Query-LIFE.} We replicated Query-LIFE  \cite{Zhu2023QueryLIFEQL} in the short video advertising scenario. It is a multimodal model based on a triple-stream architecture, where the query, video text, and video image are independently encoded before fusion. Additionally, it requires two stages of pre-training. The first stage conducts contrastive learning pre-training on video text-image pairs. The second stage utilizes online click data of <query, video> for pre-training. This click data comprises 60M query-video pairs, derived from advertisement delivery data with a click-through rate exceeding 0.2 and at least 5 impressions over the past year. Note that the Query-LIFE we replicated does not include the GenFilt component, as GenFilt is a general trick that can also be effective for other methods.

\textbf{Qwen-VL.} We fine-tuned Qwen-VL \cite{Bai2023QwenVLAV} using the same training objectives as HCMRM. Qwen-VL is a multimodal large language model that supports Chinese. Since Qwen-VL is a decoder-only architecture, the output embedding of the last token of the input sequence is used as the global multimodal representation for relevance estimation. In contrast, the above methods use the first token of the input sequence, i.e., \verb|[CLS]|.

\textbf{MiniCPM-V-2.6.} We fine-tuned MiniCPM-V-2.6 \cite{Yao2024MiniCPMVAG} using the same approach as Qwen-VL. MiniCPM-V-2.6 is also a popular multimodal large language model that supports Chinese.

\subsection{Experimental Setup} 
The number of images extracted from a short video was set to 3.
The maximum sequence lengths for the video text and query were set to 64 and 16, respectively. For pre-training, we trained HCMRM and baselines (excluding Qwen-VL and MiniCPM-V-2.6) for 3 epochs with a batch size of 256 and a learning rate of 5e-5 on 8 NVIDIA V100 GPUs. For fine-tuning, we trained HCMRM and the baselines (again excluding Qwen-VL and MiniCPM-V-2.6) on 8 NVIDIA V100 GPUs, while Qwen-VL and MiniCPM-V-2.6 were trained on 8 NVIDIA A800 GPUs. All models were trained for 4 epochs with a batch size of 64. The learning rate for Qwen-VL and MiniCPM-V-2.6 was set to 5e-7, while the learning rate for the other models was 5e-6. For learning rate scheduling, we employed a cosine decay strategy and a linear warmup of 2000 steps. We utilized the AdamW optimizer with $\beta_1 = 0.9$, $\beta_2 = 0.999$, and a weight
decay rate of 0.02.

\begin{table*}[t]
\centering
\caption{Offline experimental results compared with different baselines on the short video relevance matching task.}
 \resizebox{0.8\textwidth}{!}{ 
\begin{tabular}{llccccc}
\toprule
 \textbf{Modality} & \textbf{Model} & \textbf{X-stream} & \textbf{Query-Aware} & \textbf{AUC} & \textbf{Spearman} & \textbf{Pearson} \\
 \midrule
 \multirow{2}{*}{unimodal}
& BERT  & single & \ding{55} & 0.864 & 0.733 & 0.743   \\
& BERT w/ PQVM & single & \ding{51} & 0.868 & 0.742 & 0.752 \\  
\midrule
\multirow{6}{*}{multimodal}
& BERT-VL & single & \ding{55} & 0.869 & 0.741 & 0.750  \\
& BERT-VL w/ PQVM & single & \ding{51} & 0.872 & 0.748 & 0.758  \\  
\cmidrule(r){2-7}
& ViLT & single  & \ding{55} &  0.871 & 0.746 &  0.755 \\ 
& ViLT w/ PQVM & single & \ding{51} &  0.874 &  0.753 & 0.761  \\  
\cmidrule(r){2-7}
& Query-LIFE & triple  & \ding{51} &  0.868 & 0.742 & 0.751  \\
& Query-LIFE w/ MLM & triple & \ding{51} &  0.872 & 0.747 & 0.756  \\
\cmidrule(r){2-7}
& ALBEF  & dual & \ding{55} & 0.875  & 0.755 &  0.765 \\  
& ALBEF w/ PQVM (HCMRM) & dual & \ding{51} & \textbf{0.878} & \textbf{0.760} &  \textbf{0.770}  \\  
\bottomrule
\multicolumn{7}{l}{\footnotesize "X-stream" refers to the architecture. E.g., ALBEF is dual-stream as it uses two encoders (text and image) to process the inputs before fusion.} \\
\multicolumn{7}{l}{\footnotesize "Query-Aware" refers to whether a model implements query awareness during pre-training.}\\
\end{tabular}}
\label{mian_exp}
\end{table*}

\subsection{Offline Experimental Results} 

\subsubsection{Relevance Model.} We compared HCMRM with baselines on the short video relevance matching task. These baselines involve text-only unimodal models (BERT), single-stream multimodal models (BERT-VL, ViLT), dual-stream multimodal models (ALBEF), and triple-stream multimodal models (Query-LIFE). We also implemented the pseudo-query-video matching mechanism (PQVM) on baseline methods to verify the effectiveness of the proposed method. All models were pre-trained on domain data of short video ads, and they utilized the same relevance loss of hierarchical softmax for fine-tuning. The experimental results are shown in Table~\ref{mian_exp}.

\textbf{HCMRM achieves the best performance in short video relevance tasks.} 
Compared to baselines BERT, BERT-VL, ViLT, Query-LIFE, and ALBEF, the proposed HCMRM achieves relative improvements in the AUC score by 1.6\%, 1.0\%, 0.8\%, 1.2\%, and 0.3\%, respectively. In terms of Spearman's rank correlation coefficient, it achieves relative improvements of 3.7\%, 2.6\%, 1.9\%, 2.4\%, and 0.7\%, respectively. The changes in the Pearson Product-Moment Correlation Coefficient metric are largely consistent with those in the Spearman metric. 
In addition, multimodal methods significantly outperform text-only methods. For instance, the most lightweight multimodal method, BERT-VL, surpasses BERT by 0.6\%, 1.1\%, and 0.9\% in the AUC, Spearman, and Pearson metrics, highlighting the importance of multimodal modeling in short video scenario.
Among multimodal methods, the dual-stream method ALBEF outperforms the single-stream methods BERT-VL and ViLT, achieving increases of 0.7\% and 0.5\% in AUC scores, respectively, and enhancements of 1.9\% and 1.2\% in Spearman scores. This may be attributed to the lack of alignment between the visual and textual modalities in the single-stream approaches compared to dual-stream ones. Additionally, the performance of Query-LIFE (triple-stream) is only comparable to that of BERT-VL in short video relevance tasks. By adding the masked language modeling pre-training task, the performance of Query-LIFE has been significantly enhanced. However, there remains a noticeable performance gap compared to HCMRM.

\textbf{The pseudo-query-video matching mechanism can enhance the performance of various baselines.} HCMRM utilizes the PQVM mechanism during pre-training, achieving 0.3\% and 0.7\% improvement in AUC and Spearman metrics, compared to ALBEF. Additionally, the BERT, BERT-VL, and ViLT models that incorporate the PQVM mechanism achieve relative improvements of 0.5\%, 0.3\%, and 0.3\% in the AUC score, respectively, compared to their standard versions. In terms of Spearman's rank correlation coefficient, the relative improvements are 1.2\%, 0.9\%, and 0.9\%, respectively. These results validate the effectiveness of the pseudo-query-video matching mechanism in enhancing the query-video relevance task.

\begin{table}[h]
\centering
\caption{Offline experimental results with different relevance losses.}
\resizebox{0.85\linewidth}{!}{
\begin{tabular}{lccc}
\toprule
\textbf{Relevance loss} & \textbf{AUC} & \textbf{Spearman} & \textbf{Pearson} \\
 \midrule
binary cross-entropy &  \textbf{0.884} & 0.668 & 0.685 \\  
mean squared error &  0.876 &  0.756 & 0.766  \\
ordinal regression &  0.876 & 0.758 & 0.768  \\  
hierarchical softmax & 0.878 & \textbf{0.760} &  \textbf{0.770}  \\ 
\bottomrule
\end{tabular}}
\label{mian_exp_loss}
\end{table}

\subsubsection{Relevance Loss.} We compared HCMRM combined with different relevance fine-tuning losses. Binary cross-entropy loss, where Bad and Less are considered as 0 and Good and Excellent as 1, is employed in our previous relevance models. Mean squared error loss aims to perform regression prediction of the labels 0, 1, 2, and 3. Ordinal regression \cite{Ye2023QueryawareMB} is a competitive method that accounts for the ordinal information within relevance labels.

As shown in Table~\ref{mian_exp_loss}, the proposed relevance loss of hierarchical softmax achieves the highest scores in both Spearman and Pearson metrics.
Although hierarchical softmax loss is not as good as binary cross-entropy loss in terms of the AUC metric ($-0.7\%$), it far surpasses it in Spearman ($+13.8\%$) and Pearson metrics ($+12.4\%$).
This indicates that binary cross-entropy loss has a strong discriminative ability but weak ranking capability. Compared to mean squared error and ordinal regression loss, hierarchical softmax outperforms across all three metrics: AUC, Spearman, and Pearson, ultimately leading to a stronger overall capability.

\subsection{Online Experimental Results}

\subsubsection{Relevance Model.}

We conducted online A/B tests to compare HCMRM with our previous baseline, BERT-VL. Both models were fine-tuned using the binary cross-entropy loss. Details of the model's online deployment can be found in Appendix~\ref{serving}.
For the metric of Irrelevant Ratio, we invited annotators to evaluate the relevance performance of both models. Specifically, 10000 query-short video ad pairs were randomly sampled from the experimental buckets of both HCMRM and BERT-VL for relevance annotation. We report the absolute difference in the Irrelevant Ratio between the two test sets. For \#Conversion and Ad Revenue, we report the change of HCMRM relative to BERT-VL. As shown in Table~\ref{abtest_evaluation}, HCMRM achieves a 4\% absolute reduction in the Irrelevant Rate. As ads are more aligned with search intents, HCMRM outperforms BERT-VL by improving the number of advertising conversions (\#Conversions +1.5\%), which further led to a 1\% increase in ad revenue. 

\begin{table}[h]
\centering
\caption{Online experimental results of HCMRM compared to BERT-VL.}
\resizebox{0.8\linewidth}{!}{
\begin{tabular}{lccc}
\toprule
 & \textbf{Irrelevant Ratio} & \textbf{\#Conversion} & \textbf{Ad Revenue}  \\
 \midrule
$\Delta$ &  -4.0\%  & +1.5\% & +1.0\% \\  
\bottomrule
\end{tabular}}
\label{abtest_evaluation}
\end{table}

\subsubsection{Relevance Loss.}

We conducted online A/B tests to compare the proposed relevance loss of hierarchical softmax with the previous relevance loss of binary cross-entropy. Both sets of experiments utilized the HCMRM model. As shown in Table~\ref{abtest_evaluation_1}, the proposed relevance loss achieves a $2.1\%$ reduction in Irrelevant Rate, a $0.9\%$ increase in \#Conversion, and a $0.4\%$ improvement in ad revenue.

\begin{table}[h]
\centering
\caption{Online experimental results of hierarchical softmax relevance loss compared to binary cross-entropy loss.}
\resizebox{0.8\linewidth}{!}{
\begin{tabular}{lccc}
\toprule
 & \textbf{Irrelevant Ratio} & \textbf{\#Conversion} & \textbf{Ad Revenue}  \\
 \midrule
$\Delta$ &  -2.1\%  & +0.9\% & +0.4\% \\  
\bottomrule
\end{tabular}}
\label{abtest_evaluation_1}
\end{table}

\subsection{Advanced Topics}

\subsubsection{How do pseudo queries surpass real queries ?}
Related literature \cite{Zhu2023QueryLIFEQL, Ye2023QueryawareMB} primarily uses click data for pre-training to achieve query awareness, thereby improving consistency between pre-training and downstream relevance tasks. We collected a dataset of 60M query-video click instances (with a click-through rate > 0.2 and at least 5 impressions) for the query-aware pre-training of ALBEF. The dataset is also utilized for the Query-LIFE baseline, as mentioned above.

\begin{table}[h]
\centering
\caption{Experimental results compared with ALBEFs pre-trained on the click data.}
\resizebox{1.0\linewidth}{!}{
\begin{tabular}{llcccc}
\toprule
\textbf{Id} & \textbf{Method} & \textbf{Pretrain Data} & \textbf{AUC} & \textbf{Spearman} & \textbf{Pearson}  \\
 \midrule
1 & ALBEF & material & 0.875  & 0.755 &  0.765 \\
2 & ALBEF w/ QVM & click & 0.869 & 0.743 &  0.753 \\
3 & 1 then 2 & material + click & 0.876 & 0.757 & 0.766 \\
\midrule
4 & HCMRM  & material & \textbf{0.878} & \textbf{0.760} &  \textbf{0.770}\\
\bottomrule
\end{tabular}}
\label{mian_exp_query}
\end{table}

As shown in Table~\ref{mian_exp_query}, pre-training ALBEF on 60M click data with an additional query-video matching task (2nd row) performs noticeably worse than pre-training it on 200M video materials (1st row). This difference can be attributed to two factors. First, click data is inherently noisy, and a high click-through rate does not necessarily indicate high relevance due to exposure bias and clickbait \cite{Zou2021PretrainedLM}. Second, given the long-tail distribution, most videos receive few impressions or clicks, leading to a limited amount of query-video click data available for effective query-aware pre-training.

Furthermore, we implemented a two-stage pre-training for ALBEF: initially training on 200M video materials, followed by training on 60M click data, incorporating an additional query-photo matching task. The offline metrics of this two-stage pre-trained ALBEF are slightly better than those of the ALBEF pre-trained on 200M videos (3rd row vs. 1st row). However, it still underperforms compared to HCMRM (3rd row vs. 4th row). This indicates that although the pseudo-query mechanism of HCMRM is simple, it is effective.

\subsubsection{How does HCMRM compare to multimodal large language models?}

Multimodal large language models (MLLMs) have recently advanced significantly. We conducted a comparative evaluation of the proposed HCMRM against MLLMs on short video relevance tasks. We fine-tuned two MLLMs that support Chinese, Qwen-VL \cite{Bai2023QwenVLAV} and MiniCPM-V-2.6 \cite{Yao2024MiniCPMVAG}.

\begin{table}[h]
\centering
\caption{Experimental results compared with multimodal large language models.}
\resizebox{0.85\linewidth}{!}{
\begin{tabular}{lccccc}
\toprule
\textbf{Model} & \textbf{MLLM} & \textbf{AUC} & \textbf{Spearman} & \textbf{Pearson}  \\
 \midrule
Qwen-VL & \ding{51} & 0.872 & 0.754 &  0.765 \\
MiniCPM-V-2.6 & \ding{51} & 0.871 & 0.754 &  0.766 \\  
HCMRM & \ding{55} & \textbf{0.878} & \textbf{0.760} &  \textbf{0.770} \\ 
\bottomrule
\end{tabular}}
\label{mian_exp_llm}
\end{table}

As shown in Table~\ref{mian_exp_llm}, MLLMs do not outperform HCMRM on short video relevance tasks. In contrast, HCMRM outperforms Qwen-VL and MiniCPM-V-2.6 by 0.7\% / 0.8\% in AUC scores and 0.8\% / 0.8\% in Spearman scores. This is mainly because HCMRM has been pre-trained on hundreds of millions of short videos, whereas MLLMs lack such domain-specific knowledge. However, MLLMs still outperform multimodal methods like BERT-VL and ViLT which have been pre-trained on domain data, as shown in Table~\ref{mian_exp}. This highlights the immense potential of MLLMs for domain adaptation training when computational resources are not a constraint.

\section{Conclusion}
Mainstream vision-language pre-training methods, such as ALBEF \cite{Li2021AlignBF} and BLIP \cite{Li2022BLIPBL}, have a limited contribution to the relevance task of search ads due to the lack of query awareness. Existing query-aware multimodal methods \cite{Zhu2023QueryLIFEQL, Ye2023QueryawareMB} usually rely on noisy and sparse click data, and the introduction of an additional query tower increases the model's complexity. To address these limitations, we propose HCMRM by building upon ALBEF without significant modification costs. HCMRM enhances the consistency between multimodal pre-training and downstream relevance tasks using a simple pseudo-query-video matching mechanism. Furthermore, the introduction of hierarchical softmax loss enhances the ranking capability of relevance models, optimizing the ranking of ad systems. Ultimately, the proposed method achieves significant improvements to the Kuaishou search advertising system, with a 6.1\% reduction in the rate of irrelevant ads and a 1.4\% increase in ad revenue.


\bibliographystyle{ACM-Reference-Format}
\balance
\bibliography{sigconf}

\appendix

\section{Online Serving}
\label{serving}
To meet the real-time requirements of the production environment, the relevance model we deploy online is a lightweight multimodal network. This model is based on the BERT-VL architecture introduced in Section~\ref{baselines}. In contrast, it has fewer Transformer layers and a smaller hidden layer dimension. Additionally, it features a larger vocabulary to reduce the length of the input sequences. The visual embeddings generated by MobileNetV2 are precomputed and cached, rather than being calculated in real-time, before they are fed into BERT-VL.

Based on knowledge distillation, an offline pre-trained and fine-tuned relevance model serves as the teacher, while an online lightweight model acts as the student, facilitating the transfer of relevance matching capabilities. Specifically, this is achieved by enabling the student model to learn the relevance scores generated by the teacher model across hundreds of millions of randomly sampled query-video ad pairs.

\end{document}